\documentclass[journal]{IEEEtran}

\usepackage{xcolor,soul,framed} 
\usepackage{amsfonts}
\usepackage{float}
\usepackage[ruled,vlined,linesnumbered]{algorithm2e}

\usepackage{biblatex}
\colorlet{shadecolor}{yellow}
\usepackage[pdftex]{graphicx}
\graphicspath{{../pdf/}{../jpeg/}}
\DeclareGraphicsExtensions{.pdf,.jpeg,.png}

\usepackage[cmex10]{amsmath}
\usepackage{array}
\usepackage{mdwmath}
\usepackage{mdwtab}
\usepackage{eqparbox}
\usepackage{url}

\DeclareUnicodeCharacter{2217}{ }
\begin{document}
\bstctlcite{IEEEexample:BSTcontrol}
    \title{Application of WGAN-GP in recommendation and Questioning the relevance of GAN-based approaches}
  \author{Hichem~Ammar~Khodja, Oussama~Boudjeniba,\\~\IEEEmembership{University of Science and Technology Houari Boumediene}\\
  }

\markboth{}{}

\maketitle

\begin{abstract}
Many neural-based recommender systems were proposed in recent years and part of them used Generative Adversarial Networks (GAN) to model user-item interactions. However, the exploration of Wasserstein GAN with Gradient Penalty (WGAN-GP) on recommendation has received  relatively less scrutiny. In this paper, we focus on two questions : 1- Can we successfully apply WGAN-GP on recommendation and does this approach give an advantage compared to the best GAN models? 2- Are GAN-based recommender systems relevant? To answer the first question, we propose a recommender system based on WGAN-GP called CFWGAN-GP which is founded on a previous model (CFGAN). We successfully applied our method on real-world datasets on the top-k recommendation task and the empirical results show that it is competitive with state-of-the-art GAN approaches, but we found no evidence of significant advantage of using WGAN-GP instead of the original GAN, at least from the accuracy point of view. As for the second question, we conduct a simple experiment in which we show that a well-tuned conceptually simpler method outperforms GAN-based models by a considerable margin, questioning the use of such models.
\end{abstract}


%
\IEEEpeerreviewmaketitle


\section{Introduction}

\IEEEPARstart{I}{n} the Internet, where the number of choices is overwhelming, there is a need to filter, prioritize and effectively deliver relevant information in order to alleviate the problem of information overload, which has created a potential problem for many Internet users. Recommender systems solve this problem by searching through a large volume of dynamically generated information to provide users with personalized content and services. They have the ability to predict whether a particular user prefers an item or not based on their profile. These systems are now part of our everyday life, and in fact, almost all of our interactions with the Internet, at one point or another, involve a recommender system.

Optimizing the performance of recommendation systems comes down, among other things, to reduce the effort that users put into finding items they like, helping them make better decisions in their online purchases, increasing the number of sales, and retain customers. The fields of application are numerous and diverse (online purchase, recommendation of videos, photos, music and scientific articles, search engines, etc.) which emphasizes the potential of this technology.

Several approaches have been proposed to solve this problem, among the most popular is the model-based approach, which uses models to represent User-Item interactions. Among them are classical methods based on matrix factorization and, more recently, methods based on deep learning.

Generative Adversarial Networks (GAN) \cite{goodfellow2014generative} are a recent innovation in the field of Deep Learning. Supposing that a set of points (the training set) are sampled from a probability distribution, a GAN can approximate this distribution and effectively sample new points from it which makes the GAN a perfect tool for data generation tasks such as image generation. Despite its excellent results, it has gained some notoriety among researchers for its instability during the training process.

Wasserstein GAN with Gradient Penalty (WGAN-GP) \cite{wgan} \cite{wgan_gp} is a variant of the original GAN that partially solves the instability issue. WGAN-GP minimizes a different loss function called, the Wasserstein distance which provides more significant gradients and more stable learning.

Although GANs have been used in recommender systems, we note that the  exploration of Wasserstein GAN with Gradient Penalty (WGAN-GP) on recommendation has received relatively less scrutiny.

In this paper, we focus on two questions : 
\begin{enumerate}
    \item Can we successfully apply WGAN-GP on recommendation and does this approach give an advantage compared to the best GAN models?
    \item Are GAN-based recommender systems relevant?
\end{enumerate}
To answer the first question, we propose another approach to the GAN-based recommendation problem, which consists in modeling implicit User-Item interactions using a WGAN-GP architecture. Our model, called CFWGAN-GP, is based on CFGAN \cite{cfgan} which is a GAN-based recommender system that we adapted to our work to include WGAN-GP. A user' preferences are represented as a vector where each elements of the vector quantifies the preference of the user for a particular item, this vector is called the interaction vector. Our model is trained to generate a realistic interaction vector given information that characterizes the concerned user's profile. We successfully applied our method on two real-world datasets (MovieLens-100K and MovieLens-1M) on the top-k recommendation task and the empirical results show that it is competitive with state-of-the-art GAN approaches, but we found no evidence of significant advantage of using WGAN-GP instead of the original GAN, at least from the accuracy point of view.

As for the second question, we conduct a simple experiment in which we show that a well-tuned conceptually simpler method outperforms GAN-based models by a considerable margin, questioning the use of such models.

The main contributions of this paper are summarized as follows:
\begin{itemize}
    \item We propose a recommender system that uses Wasserstein GAN with Gradient Penalty to model User-Item interactions.
    
    \item We conduct experiments on two real-world datasets to show the effectiveness of the proposed approach compared to the best GAN models.
    
    \item We found no evidence of significant advantage of using WGAN-GP instead of the original GAN, at least from the accuracy point of view.
    
    \item We find that a well-tuned simpler method outperforms GAN models, questioning the relevance of such models in recommendation.
\end{itemize}

The rest of the paper is organized as follows : In Section \ref{section2}, we explain the preliminaries of our work and review relevant prior work on GAN-based recommender systems. In Section \ref{section3}, we design CFWGAN-GP and detail how it works. In Section \ref{section4}, we conduct experiments in order to answer the questions discussed previously. Finally, in Section \ref{section5}, we summarize our findings and results.

\section{Related work and Preliminaries}
\label{section2}
In this section, we will briefly introduce the principles of WGAN-GP and then present CFGAN, the method on which our approach is based.

\subsection{Generative Adversarial Networks (GAN)}
A GAN consists of two neural networks which are called: the generator $G$ and the discriminator $D$. Given a database, the purpose of the discriminator is to decide whether a data point comes from this base or not, the generator, for its part, aims to deceive the discriminator by generating data points as similar as possible to those present in the database. These two neural networks play a two-player minimax game: $G$ tries to deceive $D$ which increases its score but decreases that of $D$ while $D$ attempts to differentiate the real data points from the fake ones as well as possible, which increases its score but decreases that of $G$. This translates mathematically to:
\begin{equation}
\begin{split}
    \min_{G}\max_{D} f(D, G) = \mathbb{E}_{x \sim p_{data}}[log(D(x))] + \\ \mathbb{E}_{z \sim p_{noise}}[log(1 - D(G(z)))]
\end{split}
\end{equation}
where $p_{data}$ is the real data distribution and $p_{noise}$ is the distribution of the random vector. A random vector is used to add diversity to data points produced by the generator, without a random vector, a generator would return the same data point. The discriminator returns a scalar in (0, 1) which represents the probability that the data point comes from $p_{data}$.

To train the GAN, we alternatively optimize $f(D, G)$ with respect to the generator $G$ and the discriminator $D$ using a stochastic gradient descent algorithm. After enough iterations, the generator should produce data points that are close to the data points present in the dataset. Although GANs perform very well, they have gained some notoriety because of the instability of the loss function during the training phase along with the difficulty of its convergence. 

\subsection{Wasserstein GAN with Gradient Penalty (WGAN-GP)}
To alleviate the learning instabilities of the GAN, WGAN-GP was introduced. While the original GAN optimizes the KL divergence or JS divergence between the real distribution $p_r$ and generator distribution $p_g$, WGAN-GP optimizes the Wasserstein distance between $p_r$ and $p_g$ which provides better gradients for the training phase and more stable convergence. The Wasserstein distance can be defined as follows:
\begin{equation}
    W(p_r, p_g) = \sup_{\Vert f\Vert_L \leq 1} \underbrace{\mathbb{E}_{x \sim p_r}\left[f(x)\right] - \mathbb{E}_{x \sim p_g} \left[f(x)\right]}_{C}
\end{equation}
where $\Vert f\Vert_L \leq 1$ means that $f$ must be a 1-Lipschitz function.

Wasserstein GAN works as follows : We first estimate $W(p_r, p_g)$ by maximizing the term $C$ with the discriminator\footnote{In the context of Wasserstein GAN, the discriminator is called the "critic". We kept the same name so as not to overload the terminology.} and then, we minimize this approximation using the generator in order to reconcile the actual distribution of the data $p_r$ with the generated distribution $p_g$. This translates mathematically to :
\begin{equation}
\label{eq:minmax_wgan}
    \min_{G}\max_{D \in \mathcal{D}} f(G, D) = \mathbb{E}_{x \sim p_r}\left[D(x)\right] - \mathbb{E}_{z \sim p_{noise}} \left[D(G(z))\right]
\end{equation}
where $\mathcal{D}$ is the set of 1-Lipschitz functions.

In this article \cite{wgan_gp}, the authors proved that the optimal solution for the discriminator $D^*$ in Equation \ref{eq:minmax_wgan} given a fixed generator, has a gradient norm equal to 1 almost everywhere under $p_r$ and $p_g$. So to ensure the Lipschitz constraint, the authors added a term, that is called \textbf{Gradient Penalty} to the loss function of the discriminator that penalizes it when its gradient differs too much from 1. We derive the loss functions of the discriminator $J^D$ and the generator $J^G$ as follows:
\begin{equation}
\label{eq:wgan_gp}
\begin{split}
    J^D = -\mathbb{E}_{x \sim p_r}\left[D(x)\right] + \mathbb{E}_{z \sim p_{noise}} \left[D(G(z))\right] + \\\underbrace{\lambda\ \mathbb{E}_{\hat{x} \sim p_{\hat{x}}}\left[\Vert\nabla_{\hat{x}} D(\hat{x})\Vert_2 - 1\right]^2}_{\textrm{Gradient Penalty}}
\end{split}
\end{equation}
\begin{equation}
    J^G = -\mathbb{E}_{z \sim p_{noise}} \left[D(G(z))\right]
\end{equation}
where $\lambda$ controls the intensity of the gradient penalty. Sampling from $p_{\hat{x}}$ is done by taking random points in the segment between fake and real data points.

As in the original GAN, the training of the WGAN-GP is done by alternating between the optimization of $J^G$ and $J^D$ using a stochastic gradient descent algorithm.

\subsection{Recommender systems}
Collaborative Filtering (CF) uses algorithms to learn patterns from user-item interactions to make personalized recommendations for users. Among CF methods, model-based methods use models to represent the interactions between users and items. Some of the most popular model-based algorithms use matrix factorization which learns the linear interactions between latent features of users and items (BiasedMF \cite{biasedmf}, SVD++ \cite{svd++}, SLIM\cite{slim}). Due to the huge success of deep learning, neural-based approaches were proposed to discover non-linear user-item interactions (NCF \cite{ncf}, CDAE \cite{cdae}).

More recently, GAN-based recommender systems were introduced, the first ones being IRGAN \cite{irgan} and GraphGAN \cite{graphgan}. These two methods are quite similar so we only briefly present the former. In IRGAN, the generator, given a user's profile, tries to generate single items that are relevant to the user, and the discriminator tries to discriminate the user's ground truth items from the ones produced by the generator. At first, the generator produces random items but it will improve gradually with the guidance of the discriminator. At one point, the generator will produce items identical to those in the ground truth which makes the task of the discriminator quite challenging because can receive the same item which is sometimes labeled as "fake" and sometimes labeled as "real", leading to the degradation of the learning process.

\subsection{CFGAN}
The authors of CFGAN start by identifying the fundamental problem of IRGAN and GraphGAN, discussed earlier, and then they propose a new GAN-based framework to solve the problem. The GAN is conditioned on the user \cite{mirza2014conditional} and it uses a vector-wise adversarial approach with implicit feedback: Instead of producing single items (like IRGAN), the generator produces the vector of preferences of all items and the discriminator has to differentiate between fake and real vectors. CFGAN uses binary implicit feedback so the generator's output and the ground truth are in $[0, 1]^n$ where $n$ is the number of items.

Given a user-specific condition vector $c_u$, the CFGAN generates an $n$-dimensional vector $\hat{x}_u$ where $\hat{x}_{u_i}$ is hopefully close to 1 if the user $u$ interacted with item $i$ ($i \in \{1, 2, ..., n\}$). Formally, both the discriminator and the generator's loss functions, denoted as $J^D$ and $J^G$ respectively, are set as follows:
\begin{equation}
\label{eq:jd_cfgan}
    J^D = -\cfrac{1}{|U|} \sum_{u \in U} \bigg[ log(D(x_u | c_u)) + log(1-D(\hat{x}_u \odot x_u | c_u)) \bigg]
\end{equation}
\begin{equation}
\label{eq:jg_cfgan}
    J^G = \cfrac{1}{|U|}\sum_{u \in U}log(1 - D(\hat{x}_u \odot x_u | c_u))
\end{equation}
where $U$ is a batch of users, $x_u$ is the ground truth interaction vector of the user $u$, $\hat{x}_u$ is the interaction vector produced by the generator, and $\odot$ denotes an element-wise multiplication. $D(x_u|c_u)$ represents the probability that the vector $x_u$ is real given the condition vector $c_u$.

The difference between the original GAN and CFGAN is that first, random vectors are not used in this architecture since the goal is to generate the single, most plausible recommendation result to the target user rather than multiple outcomes. Second, before feeding the generator's output to the discriminator, we first multiply it by the ground truth interaction vector $x_u$ which results in masking the items in $\hat{x}_u$ that the user $u$ did not interact with. This is done to deal with the sparsity in ground-truth data that the generator tries to mimic. By doing so, only the generator’s output on the interacted items can contribute to the learning of the generator and the discriminator.

Both the discriminator $D$ and the generator $G$ are implemented as multi-layer neural networks. $G$ takes as input a user-specific conditional vector $c_u$ and outputs the presumed interaction vector $\hat{x}_u$. $D$ takes as input $c_u$ concatenated with $x_u$ (real data) or $c_u$ concatenated with $\hat{x}_u \odot x_u$ (fake data) and returns a scalar value representing the probability that the input comes from the ground truth. $J^D$ and $J^G$ are alternatively minimized using a stochastic gradient descent algorithm.

This architecture, as it is, will not work because the generator will arrive at the trivial solution of returning a vector with all elements equal to 1. After all, if we compute the element-wise multiplication of this vector with the ground truth vector $x_u$, the result will be $x_u$, and the discriminator will have no clue how to differentiate between real and fake data, and in the end, the generator just learned to produce full-of-ones vectors which are completely useless.

The solution proposed by the authors is to add a penalty to the loss function of $G$ to prevent it from reaching the trivial solution. They called this design \textit{CFGAN\_ZR} (\textit{ZR} means "Zero Reconstruction") and it works as follows: For each user $u$ used to train $G$, we select random items that did not interact with $u$\footnote{When we use the expression "item $i$ interacted with user $u$" or vice versa, it means that there is an implicit feedback involving $u$ and $i$.} and we apply a penalty to the generator when the output for those particular items differ from 0.

The authors added another modification to the loss function called \textit{CFGAN\_PM} (\textit{PM} means "Partial Masking") and it works as follows: Instead of masking every item that did not interact with the user in the generator's output $\hat{x}_u$, we mask only a portion of it and we control the size of this portion with a parameter. Consequently, the discriminator now exploits the input values from not only the interacted items but also from the negative ones\footnote{If a user does not have an interaction with a particular item, it is called a \textit{negative item} of that user.}. Moreover, the gradient of the loss from the discriminator with respect to the output on the selected negative items can be passed back to the generator to better guide its learning.

When we combine these two designs we get what the authors called \textit{CFGAN\_ZP}. We will come back to these designs in more detail in Section \ref{section3}.

\section{CFWGAN-GP : Recommender system based on WGAN-GP}
\label{section3}
\begin{figure}
  \begin{center}
  \includegraphics[width=3.5in]{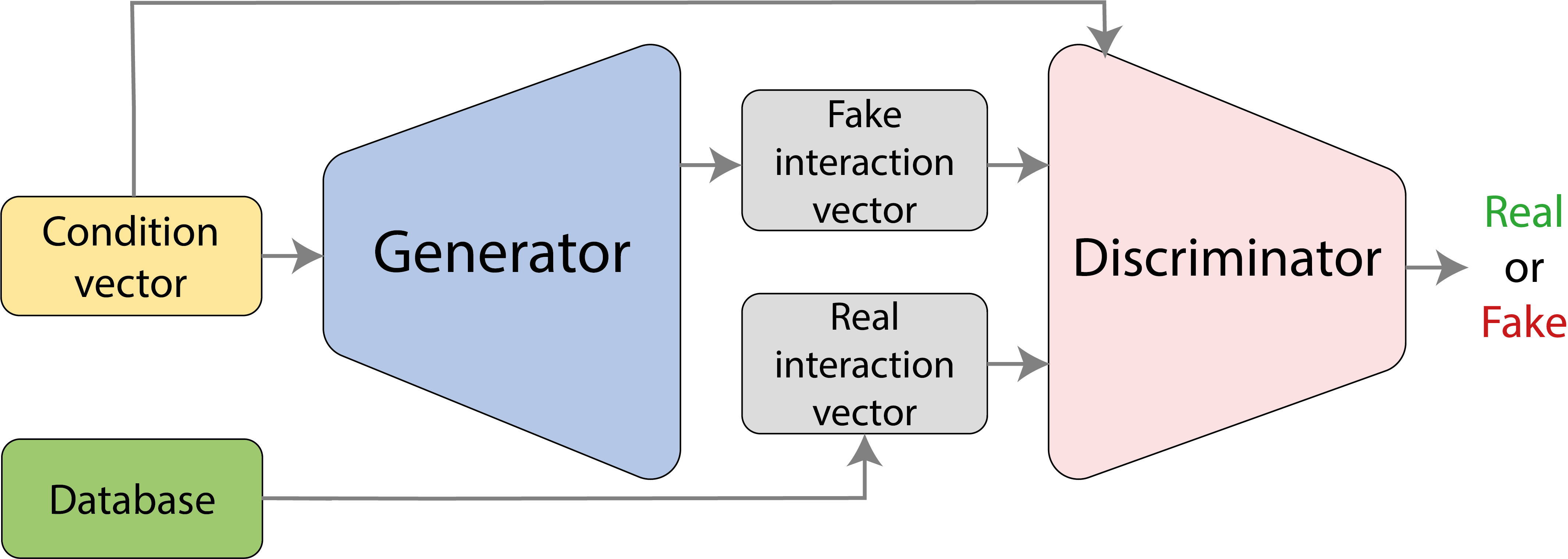}\\
  \caption{Overview of our recommender system (CFWGAN-GP)}\label{fig:cfwgan_gp}
  \end{center}
\end{figure}

In this part, it is a question of carrying out an approach based on generative antagonist networks (GAN), more precisely, by using the variant WGAN-GP. The Figure \ref{fig:cfwgan_gp} illustrates how our model works. The purpose of the generator is to predict, for a given user, his interaction vector knowing the condition vector of the user $c_u$. This vector describes the profile of the user; it can contain, for example, the user's gender, occupation, age, etc. or simply his interaction vector which is the approach chosen in this paper. The purpose of the discriminator is to differentiate the real interaction vector of the user from the ones produced by the generator. Our work is based on CFGAN, which is another GAN-based recommender system, that we adapt to our study to include WGAN-GP instead of the original GAN.

Both the generator and the discriminator are multi-layer neural networks having a ReLU activation function in all layers, except for the last layer of the generator which has a sigmoid activation function because the generation task aims to generate interaction vectors which are in $[0, 1]^n$, and the last layer of the discriminator which has an identity activation function ($f(x) = x$).

In this section, we begin by adapting CFGAN to our work by translating the loss functions from the original GAN design to the WGAN-GP design. After that, we include \textit{CFGAN\_ZP} design in our model and finally, we introduce the learning algorithm.

\subsection{From GAN to WGAN-GP}
By translating the loss functions of the discriminator and the generator (Equations \ref{eq:jd_cfgan} and \ref{eq:jg_cfgan}) into WGAN-GP settings, we directly obtain the following equations:
\begin{equation}
    \label{eq:jd_cfwgan_gp}
    J^D = \cfrac{1}{|U|} \sum_{u \in U} \bigg[ D(\hat{x}_u \odot x_u | c_u) - D(x_u | c_u)\bigg] + GP
\end{equation}
\begin{equation}
    \label{eq:jg_cfwgan_gp}
    J^G = \cfrac{1}{|U|}\sum_{u \in U}-D(\hat{x}_u \odot x_u | c_u)
\end{equation}
where $D$ is a 1-Lipschitz function and $GP$ is the gradient penalty applied to $D$ (see Equation \ref{eq:wgan_gp}). The rest of the symbols have the same denotation as in Equations \ref{eq:jd_cfgan} and \ref{eq:jg_cfgan}.

As in CFGAN, if we train the model with these loss functions, the generator will arrive at the trivial solution of returning a vector full of ones. To avoid that, we implement the \textit{CFGAN\_ZR} design.

\subsection{Zero reconstruction (CFGAN\_ZR)}
This method adds a term that penalizes the generator when, given a user, the value of an item, from the generated interaction vector, is close to 1 whereas in the ground truth, it is in reality 0 (negative item). This penalty is applied for a subset of negative items sampled randomly, without repetition, from the set of all negative items.

Let $N_u$ be the set of the negative items of the user $u$:
\begin{equation*}
    N_u = \{j \in \{1,2,\dots, n\}\mid (x_u)_j = 0\}
\end{equation*}
where $(x_u)_j$ is the $j$-th element of the interaction vector $x_u$.

Let $ZR_u$ be a subset of $N_u$ that is sampled randomly. $p_{ZR}$ is the parameter that controls the size of $ZR_u$, more precisely, $p_{ZR} = \cfrac{|ZR_u|}{|N_u|} \in [0, 1]$. 

We add the penalty to the generator's loss function as a regularization term, called $ZR$ penalty:
\begin{equation}
    \label{jg_cfwgan_gp_zr}
    J^G = \cfrac{1}{|U|}\sum_{u \in U}\left[-D(\hat{x}_u \odot x_u | c_u) + \underbrace{\alpha\sum_{j \in ZR_u}((\hat{x}_u)_j)^2}_{ZR}\right]
\end{equation}
where $\alpha$ controls the intensity of $ZR$ penalty.

\subsection{Partial masking (CFGAN\_PM)}
In Equations \ref{eq:jd_cfwgan_gp} and \ref{eq:jg_cfwgan_gp}, we mask all negative items from the generated interaction vector before sending it to the discriminator. By implementing \textit{CFGAN\_PM}, only a portion of the negative items is masked. This allows the discriminator to exploiting the positive items produced by the generator as well as a portion of the negative items. Consequently, the gradient of the loss from the discriminator with respect to those negative items can be passed to the generator for better guidance in its learning process.

Let $PM_u$ be a subset of $N_u$ that is sampled randomly. $p_{PM}$ is the parameter that controls the size of $PM_u$ ($p_{PM} = \cfrac{|PM_u|}{|N_u|}$).

Let $k_u = [k_1, k_2,\dots, k_n]$ where $k_j = 1$ if and only if $j \in PM_u$. To partially mask the negative items, instead of multiplying the generator's output by the ground truth $x_u$, we multiply it by $(x_u + k_u)$. By applying \textit{CFGAN\_PM} design to Equations \ref{eq:jd_cfwgan_gp} and \ref{eq:jg_cfwgan_gp}, we get:
\begin{equation}
    \label{eq:jd_cfwgan_gp_pm}
    J^D = \cfrac{1}{|U|} \sum_{u \in U} \bigg[ D(\hat{x}_u \odot (x_u + k_u) | c_u) - D(x_u | c_u)\bigg] + GP
\end{equation}
\begin{equation}
    \label{eq:jg_cfwgan_gp_pm}
    J^G = \cfrac{1}{|U|}\sum_{u \in U}-D(\hat{x}_u \odot (x_u + k_u)| c_u)
\end{equation}

The parameters $p_{ZR}$ and $p_{PM}$ are the same for all users and $ZR_u$ and $PM_u$ are regenerated at each learning step involving the user $u$.

\subsection{CFGAN\_ZP}
\textit{CFGAN\_ZP} is the combination of the two previous designs: We penalize the model for errors for a portion of negative items (\textit{CFGAN\_ZR}) and we only \textit{partially} mask negative items from the generator's output before feeding it to the discriminator (\textit{CFGAN\_PM}). We apply this design to the loss functions of the discriminator and the generator (Equations \ref{eq:jd_cfwgan_gp} and \ref{eq:jg_cfwgan_gp}):
\begin{equation}
    \label{eq:jd_cfwgan_gp_zp}
    J^D = \cfrac{1}{|U|} \sum_{u \in U} \bigg[ D(\hat{x}_u \odot (x_u + k_u) | c_u) - D(x_u | c_u)\bigg] + GP
\end{equation}
\begin{equation}
    \label{eq:jg_cfwgan_gp_zp}
    J^G = \cfrac{1}{|U|}\sum_{u \in U}\left[-D(\hat{x}_u \odot (x_u + k_u)| c_u) + \alpha\sum_{j \in ZR_u}((\hat{x}_u)_j)^2\right]
\end{equation}

\subsection{Learning algorithm}
To train our model, we alternatively optimize the loss functions of the discriminator and the generator using a stochastic gradient descent algorithm, being Adam\cite{kingma2017adam}, in our case. The algorithm below describes the learning process of our model. $\theta_G$ and $\theta_D$ are the parameters of the generator and the discriminator respectively and $D_{iter}$ is the number of steps in which $J^D$ must be optimized before switching to the optimization of $J^G$ one single time.

\begin{algorithm}[h]
\label{algo:cfwgan_gp}
\SetAlgoLined

\SetKwFor{RepTimes}{repeat}{times}{end}
Initialize $\theta_G$ and $\theta_D$ randomly\;
\While{\upshape not converged}{
     \RepTimes{$D_{iter}$}{
        \upshape Sample a batch of users $U$ from the dataset\;
        \upshape Generate $PM_u$ for each user $u \in U$\;
        \upshape Get real interaction vectors $x_u$ for each $u \in U$\;
        \upshape Produce the fake interaction vector $\hat{x}_u$ with the generator for each $u \in U$\;
        \upshape Compute the gradient of $J^D$ with respect to $\theta_D$\;
        \upshape Update the parameters $\theta_D$ using the Adam algorithm\;
     }
    \upshape \upshape Sample a batch of users $U$ from the dataset\;
    \upshape Generate $ZL_u$ et $PM_u$ for each user $u \in U$\;
    \upshape Produce the fake interaction vector $\hat{x}_u$ with the generator for each $u \in U$\;
    \upshape Compute the gradient of $J^G$ with respect to $\theta_G$\;
    \upshape Update the parameters $\theta_G$ using the Adam algorithm\;
}
\caption{Learning algorithm of CFWGAN-GP}
\end{algorithm}

\subsection{Prediction process}
After the model is trained, we keep only the generator for the prediction process. Let $u_{test}$ a user for whom we want to recommend the best $k$ items (top-k recommendation task). First, we get his condition vector $c_{test}$, then we produce the interaction vector with the generator $\hat{x}_{test} = G(c_{test})$ and we order the items in descending order according to the score obtained in $\hat{x}_{test}$. Next, we ignore the items that the user has already interacted with, and finally, we recommend the $k$ items with the best score to the user.

\section{Experiments}
\label{section4}
After training the model, we test the generator to measure its performance on real-world datasets. The aim of this section is to answer the following questions:
\begin{itemize}
    \item \textbf{Q1}: How much effective is our method for the top-k recommendation task compared to the best GAN models?
    \item \textbf{Q2}: Are GAN-based recommender systems relevant?
\end{itemize}

\subsection{Experimental settings}
MovieLens \cite{movielens} is a widely used film dataset for evaluating recommender systems. We used two versions: the one containing 100,000 ratings (MovieLens-100K) and the one containing one million ratings (MovieLens-1M). Each user has at least 20 ratings. This dataset initially contained explicit feedback which we converted to implicit binary feedback where each entry is marked as 1 or 0 indicating respectively whether the user has interacted with the item or not. We randomly split the user-item interactions into two subsets: A training set containing 80\% of the interactions and a test set containing the remaining 20\%. We reserve 20\% of the training set for validation in order to perform hyperparameter tuning.

\begin{figure*}
    \centering
    \includegraphics[width=\textwidth]{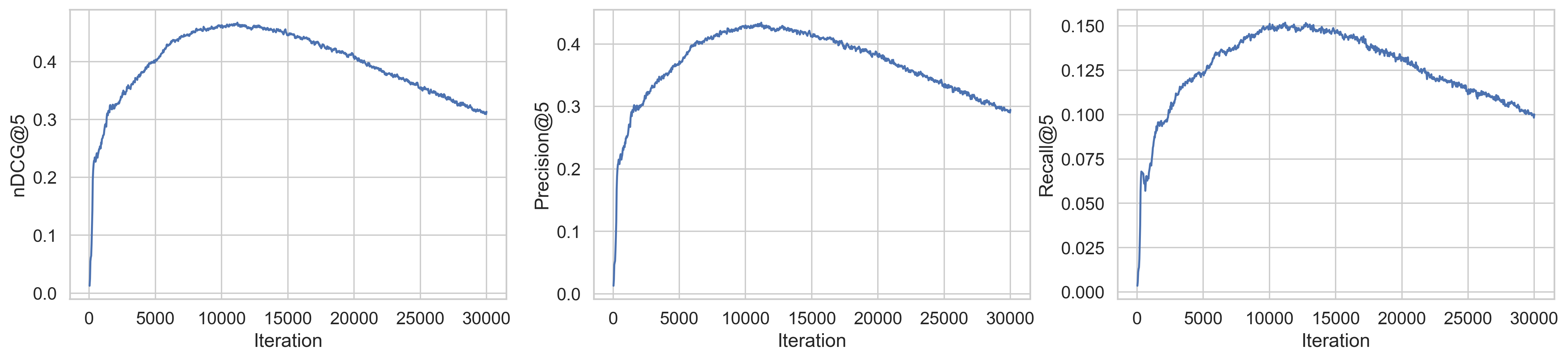}
    \caption{Learning curves of CFWGAN-GP on MovieLens-100K}
    \label{fig:curves_ml100k}
\end{figure*}
\begin{figure*}
    \centering
    \includegraphics[width=\textwidth]{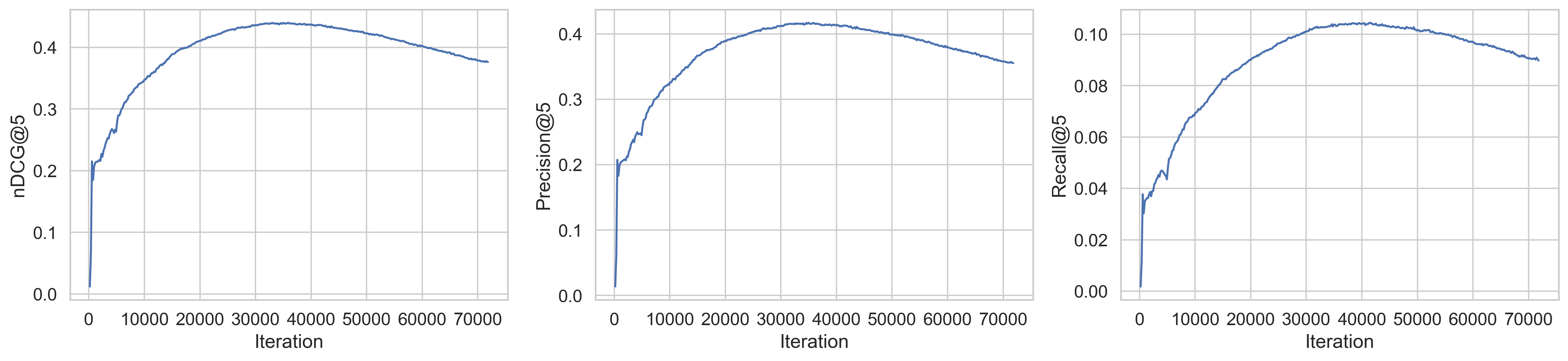}
    \caption{Learning curves of CFWGAN-GP on MovieLens-1M}
    \label{fig:curves_ml1m}
\end{figure*}

\begin{table}[h]
\center
\begin{tabular}{|c|c|c|c|c|}
\hline
\textbf{Datasets} & \textbf{\begin{tabular}[c]{@{}c@{}}Number of\\ users\end{tabular}} & \textbf{\begin{tabular}[c]{@{}c@{}}Number of\\ items\end{tabular}} & \textbf{\begin{tabular}[c]{@{}c@{}}Number of\\ ratings\end{tabular}} & \textbf{Density} \\ \hline
Movielens 100K    & 943                                                                       & 1682                                                              & 100 000                                                                   & 6.3\%            \\ \hline
Movielens 1M      & 6040                                                                      & 3706                                                              & 1 000 209                                                                 & 4.47\%           \\ \hline
\end{tabular}
\caption{General dataset statistics}
\end{table}

We use three popular metrics for top-k recommendation: Precision (P@k), Recall (R@k) and normalized discounted cumulative gain (N@k) for $k=5$ and $k=20$. The two first metrics measure the number of correct guesses that are included in the recommendation list, while the last one takes into account the order in which they came.

For hyperparameter tuning, we keep the recommended default values for $D_{iter} = 5$, $\lambda = 10$, $\beta_1 = 0$ and $\beta_2 = 0.9$\footnote{$\beta_1$ and $\beta_2$ are parameters from the Adam algorithm} from the WGAN-GP article. The parameters we are tuning are : the learning rate $lr$, the number of hidden layers for the generator $l_g$ and the discriminator $l_d$, the number of nodes per hidden layer for the generator $h_g$ and the discriminator $h_d$, the intensity of the \textit{ZR} penalty $\alpha$, $p_{ZR}$ and $p_{PM}$. Since our approaches are very similar, our starting point is the best configurations of CFGAN on MovieLens-100K and MovieLens-1M which we tune \textit{manually} to maximize N@20.

\subsection{Q1: Effectiveness of our model}
We train our model and we measure its performances according to the metrics mentioned above. 

The best performances on MovieLens-100K were obtained using the configuration ($lr=10^{-4},\ l_g=l_d=1,\ h_g=h_d=512,\ \alpha=0.04,\ p_{ZR}=0.7,\ p_{PM}=0.6$), while on MovieLens-1M, the best configuration is ($lr=10^{-4},\ l_g=l_d=1,\ h_g=300,\ h_d=250,\ \alpha=0.03,\ p_{ZR}=p_{PM}=0.7$).

Figures \ref{fig:curves_ml100k} and \ref{fig:curves_ml1m} show the learning curves of our model on MovieLens-100K and MovieLens-1M respectively. The curves of N@20, P@20, and R@20 exhibit very similar behavior to N@5, P@5, and R@5 respectively.

\begin{table*}
\centering
\resizebox{\textwidth}{!}{
\begin{tabular}{c|c|c|c|c|c|c|c|c|c|c|c|c|}
\cline{2-13}
 & \multicolumn{6}{c|}{\textbf{MovieLens-100K}} & \multicolumn{6}{c|}{\textbf{MovieLens-1M}} \\ \cline{2-13} 
 & N@5 & N@20 & P@5 & P@20 & R@5 & R@20 & N@5 & N@20 & P@5 & P@20 & R@5 & R@20 \\ \hline
\multicolumn{1}{|c|}{ItemPop} & .163 & .195 & .181 & .138 & .102 & .251 & .154 & .181 & .157 & .121 & .076 & .197 \\ \hline
\multicolumn{1}{|c|}{BPR} & .370 & .380 & .348 & .236 & .116 & .287 & .349 & .362 & .341 & .252 & .077 & .208 \\ \hline
\multicolumn{1}{|c|}{FISM} & .444 & .436 & .426 & .281 & .139 & .342 & .427 & .401 & .408 & .292 & .098 & .263 \\ \hline
\multicolumn{1}{|c|}{CDAE} & .450 & .436 & .433 & .288 & .141 & .358 & .441 & .411 & .411 & .300 & .102 & .278 \\ \hline
\multicolumn{1}{|c|}{GraphGAN} & .183 & .249 & .212 & .151 & .102 & .260 & .205 & .184 & .178 & .194 & .070 & .179 \\ \hline
\multicolumn{1}{|c|}{IRGAN} & .342 & .368 & .312 & .221 & .107 & .275 & .264 & .246 & .263 & .214 & .072 & .166 \\ \hline
\multicolumn{1}{|c|}{CFGAN} & .480 & .441 & .441 & \textbf{.302} & \textbf{.161} & .361 & .442 & .411 & .423 & \textbf{.317} & .110 & \textbf{.285} \\ \hline
\multicolumn{1}{|c|}{\textbf{Ours (CFWGAN-GP)}} & .461 & .430 & .423 & .285 & .148 & .359 & .437 & .390 & .414 & .296 & .104 & .260 \\ \hline
\multicolumn{1}{|c|}{\textbf{Ours (MLC)}} & \textbf{.486} & \textbf{.448} & \textbf{.450} & .300 & .156 & \textbf{.365} & \textbf{.472} & \textbf{.419} & \textbf{.446} & \textbf{.317} & \textbf{.111} & .280 \\ \hline
\end{tabular}}
\caption{Comparison results for MovieLens-100K and MovieLens-1M}
\label{table:perf}
\end{table*}

We now compare the accuracy of our model with the best GAN-based approaches and other top-N recommenders, which are listed as follows:

\begin{itemize}
    \item \textbf{ItemPop} : A model that always recommends the most popular items to the user (The popularity of an item is measured by the total number of interactions with users). This is one of the simplest non-personalized methods.
    \item \textbf{BPR} \cite{bpr} : This method proposes a loss function measuring the error on the order of the items of the list of recommendations that is generated by the system. This function is minimized using a variant of the stochastic gradient descent algorithm, called LearnBPR. 
    \item \textbf{FISM} \cite{fism} : This method builds an item-item similarity matrix and produces models to approximate it by factoring it similarly to matrix factorization methods. User preferences are predicted using this factorization. 
    \item \textbf{CDAE} : It uses a denoising autoencoder structure for CF
    while integrating user-specific latent features. It is trained with a user’s purchase vector as input and aims at reconstructing it as similar as possible, just like our model.
    \item \textbf{IRGAN} and \textbf{GraphGAN} : These are the first GAN-based recommender systems discussed in Section \ref{section2}.
\end{itemize}

We use the training results from the CFGAN paper to compare our model with CFGAN and the models above. We can legitimately make this comparison since we use the exact same evaluation procedure and the same datasets as in the CFGAN paper. All the models were tuned by the authors to maximize N@20.

Table \ref{table:perf} compiles the performances of all models. We note that the results of our model CFWGAN-GP are competitive with those of the best GAN models. The most important comparison to make is with the CFGAN model since our approach is based on it. Although the difference in performance between CFGAN and CFWGAN-GP can, to some extent, be explained by the insufficient tuning of hyperparameters for the latter, we found no evidence of significant improvement in accuracy after replacing the original GAN with WGAN-GP in the CFGAN model.

\subsection{Q2 : A simpler, yet more effective approach}
The GAN-based approach presented in this paper is shown to be working quite effectively ; however, was GAN necessary in the first place? The problem we are trying to solve can be formalized as a multi-label classification problem that may not require the deployment of systems as complex as the CFGAN. The aim of this experiment is therefore to determine if the GAN approach (and the other models in the table above) gives an advantage over a more classical approach like treating the problem as a multi-label classification problem and solving it with a neural network using a stochastic gradient descent algorithm.

In this experiment, we use a neural network with one hidden layer, noted $M$. The input is the condition vector of the user $c_u$ and the output is his interaction vector. We use a ReLU activation function for the hidden layer and the sigmoid activation function for the output layer. The loss function, which is the binary cross-entropy loss, is defined as follows:
\begin{equation}
    J = \frac{1}{|U|}\sum_{u \in U} mean(x_u log(\hat{x}_u) + (1-x_u) log(1-\hat{x}_u))
\end{equation}
where $U$ is a batch of users, $x_u$ is the ground truth interaction vector, $\hat{x}_u = M(c_u)$ is the predicted interaction vector and $mean(v)$ is the mean of the vector $v$.

This loss is minimized, with respect to the parameters of $M$, using the Adam algorithm with default parameters. To handle overfitting, we add dropout \cite{dropout} and L2 regularization. We manually tuned hyperparameters which are the learning rate $lr$, the L2 coefficient $\lambda$, the number of nodes of the hidden layer $h$ and the dropout rate $p$. For MovieLens-100K, we find that the configuration that works best is ($lr=10^{-4}, \lambda=10^{-5}, h=256, p=0.8$), while on MovieLens-1M, the best configuration we found is ($lr=10^{-4}, \lambda=0, h=400, p=0.8$). The training results are shown in the last row of Table \ref{table:perf}\footnote{"MLC" stands for "Multi-Label Classification".}. 

We see that a simpler method outperforms, in most metrics, all the other models including recent neural-based approaches like CFGAN and CDAE. In addition, this method, compared to CFGAN, has fewer parameters (less memory space) and its training is faster. This experiment questions the use of such models and is in line with the results of this article \cite{are_we_making_much_progress} where the authors tested many recent neural-based approaches, that were presented in top-level research conferences, and compared their accuracy with comparably simple heuristic methods, e.g., based on nearest-neighbor \cite{userKNN} \cite{itemKNN} or graph-based techniques \cite{p3} \cite{rp3}. They found that in all cases, there was at least one simple heuristic that outperforms the neural-based model that was being tested.

\section{Conclusion}
\label{section5}
In this paper, we proposed a new recommender system based on WGAN-GP called CFWGAN-GP, we studied its effectiveness compared to recommender systems that use classical GAN architectures and we achieved that by conducting several experiments on top-k recommendation task. Our model is based on the CFGAN model that we adapted to our study to include WGAN-GP, and it basically works as follows: Given a user, the generator produces a vector describing his preferences for all items, called the interaction vector, while the discriminator aims to differentiate between the fake interaction vectors and the real ones. Even though the experiments showed that our model had competitive accuracy with the best GAN models, we found no evidence of improvement in performance by using WGAN-GP architecture instead of the original GAN architecture. In the second part of the experiments, we found that the problem that CFGAN is trying to solve can be treated as a multi-label classification problem and we solved it by using a multi-layer neural network that minimizes the binary cross-entropy loss between the ground-truth interaction vectors and the ones produced by the model. This conceptually simpler method outperformed all tested models including recent neural-based approaches like CDAE and CFGAN which questions the relevance of such models. This result is in line with several recent publications' results \cite{neural_hype}\cite{eval_session_based} that point out similar problems in today’s applied machine learning in recommender systems.

\printbibliography
\end{document}